\documentclass[twocolumn,eqsecnum,aps,prd,showpacs,preprintnumbers,amsmath,amssymb,nofootinbib,floatfix]{revtex4}

\usepackage[active]{srcltx}
\usepackage{bm}
\usepackage{graphicx,epsf}
\usepackage{pstricks}
\usepackage{amssymb}

\newcommand\calp{\mathcal{P}}
\newcommand\caln{\mathcal{N}}
\newcommand\mpl{m_{\rm Pl}}
\newcommand\fnl{f_{NL}}
\renewcommand\({\left(}
\renewcommand\){\right)}
\newcommand\be{\begin{equation}}
\newcommand\ee{\end{equation}}
\newcommand\bea{\begin{eqnarray}}
\newcommand\eea{\end{eqnarray}}
\newcommand\eq[1]{equation ~(\ref{#1})}
\newcommand\fig[1]{Fig.~(\ref{#1})}

\newcommand{\Vol}{\mathrm{Vol} (X_5)}

\newcommand{\Ps}{\mathcal{P}_S^2}
\newcommand{\Pt}{\mathcal{P}_T^2}

\begin{document}
\title{Single-Field Inflation After WMAP5}
\author{Laila Alabidi}
\email{l.alabidi@qmul.ac.uk}
\author{James E. Lidsey}
\email{j.e.lidsey@qmul.ac.uk}
\affiliation{Astronomy Unit, School of Mathematical Sciences, 
Queen Mary, University of London, Mile End Road, London, E1 4NS, 
United Kingdom} 
\begin{abstract}
Single-field models of inflation are analysed in light of the  
WMAP five-year data. Assuming instantaneous reheating, 
we find that modular/new 
inflation models with small powers in the effective inflaton self-interaction
are more strongly constrained than previously. 
The model with a cubic power lies outside 
the $2\sigma$ regime when the number of e-folds 
is $\caln \le 60$. We also find that the 
predictions for the intermediate
model of inflation do not overlap the $1\sigma$ region regardless
of the power of the monomial potential.
We analyse a number of ultra-violet, 
DBI braneworld scenarios involving both wrapped and multiple-brane 
configurations, where the inflaton kinetic energy 
is close to the maximum allowed 
by the warped geometry. In all cases, we find that the parameters 
of the warped throat are strongly constrained by observations. 

\end{abstract}


\maketitle

\section{Introduction}

The inflationary scenario postulates that the universe underwent a 
phase of very rapid, accelerated expansion in its distant past. 
Observations -- most notably from the anisotropy power spectrum of the
Cosmic Microwave Background (CMB) -- have provided strong support for the 
paradigm. Despite this success, however, the mechanism which 
drove the inflationary expansion has yet to be identified. Indeed, 
there remain many viable versions of the scenario. 
(For reviews, see, e.g., Refs.  
\cite{Lyth:1998xn,Liddle:2000cg,Guth:2005zr,Cline:2006hu,HenryTye:2006uv,McAllister:2007bg,Linde:2007fr,Lyth:2007qh}.)
In view of this,
it is important to constrain, and ultimately rule out, as many models of
inflation as possible. 

In this paper, we consider a wide range of inflationary models that are well
motivated from particle physics and unified field theory. We focus on 
models where the cosmic expansion was driven by 
a single, self-interacting scalar `inflaton' field, $\phi$. In general, 
single-field inflationary models are characterised by the action 
\begin{equation}
\label{generalaction}
S= \int d^4x \sqrt{-g} \left[ \frac{\mpl^2}{2} R + P(\phi , X) \right],
\end{equation}
where $R$ is the Ricci curvature scalar, the `kinetic function' 
$P(\phi , X)$ is a function of the inflaton field and its 
kinetic energy $X \equiv - \frac{1}{2} g^{\mu\nu}\nabla_{\mu} \phi
\nabla_{\nu} \phi$, and $\mpl = (8\pi G)^{-1/2}$ is the reduced Planck mass. 
In the simplest versions of the scenario, the inflaton has a canonically
normalised kinetic energy, where  
$P=X- V(\phi )$ and $V(\phi)$ denotes the inflaton potential. 
In many higher-dimensional models motivated 
directly from string/M-theory, however, the function $P (\phi ,X)$ has a 
more complicated form. This is the case, for example, in the 
Dirac-Born-Infeld (DBI) inflationary scenario, where
inflation is driven by the propagation of one or more ${\rm D}$-branes \cite{DBI}.  

We compare both canonical and non-canonical models of inflation 
with the recent observational bounds derived from the combined data of the 
Wilkinson Microwave Anisotropy Probe (WMAP) \cite{wmap5}, 
Baryon Acoustic Oscillations (BAO) \cite{Percival:2007yw} and 
Supernovae (SN) surveys \cite{Riess:2004nr,Astier:2005qq, Riess:2006fw,WoodVasey:2007jb}. 
In Ref. \cite{Alabidi:2006qa}, the three-year WMAP data was employed 
to rule out for the first time a large fraction of viable canonical 
models at more than $3\sigma$. We begin by reconsidering 
this analysis in the light of the WMAP five-year data \cite{wmap5}. 
We find that the conclusions 
of \cite{Alabidi:2006qa} are confirmed by WMAP5, 
with the improved bounds on the scalar spectral index 
strongly constraining models of modular inflation 
at more than the 2$\sigma$ level.

We then proceed to consider non-canonical DBI brane inflationary models  
driven by a wrapped ${\rm D}5$- or ${\rm D}7$-brane \cite{DBI,BECKER,KOB}.
We focus on the `relativistic, ultra-violet' version of the scenario, 
where the brane is moving towards the tipped region of a warped throat
with a kinetic energy close to the upper bound imposed by the 
warpfactor of the higher-dimensional space. 
We find that independently of the form of the inflaton potential, 
the geometry of the warped throat must be strongly constrained 
if such models are to satisfy the improved observational bounds 
from WMAP5. We also consider extensions of the scenario 
to multiple-brane configurations \cite{thomasward,HLTW} 
and find that the same challenges 
arise as for the single-brane case.   

The structure of the paper is as follows. 
In section II, we summarise the current bounds on the observational parameters. 
In section III, we investigate models of canonical 
inflation and determine which models are still viable. 
In Sections IV and V, we derive new observational limits on wrapped and 
multi-brane UV DBI inflation. We conclude with a discussion in Section VI.  

\section{Single-Field Inflation}

For the action (\ref{generalaction}), 
the amplitudes of the scalar and tensor perturbation spectra generated 
during inflation are given by \cite{GM1999}
\begin{equation}
\label{amp}
\Ps = \frac{1}{8 \pi^2 \mpl^2}\frac{H^2}{c_s \epsilon} , \quad 
\Pt = \frac{2}{\pi^2} \frac{H^2}{\mpl^2}
\end{equation}
respectively, where $\epsilon \equiv -\dot{H}/{H^2}$, the sound speed of 
inflaton fluctuations is defined by 
\begin{equation}
\label{defcs}
c_s^2 \equiv \frac{P_{,X}}{P_{,X} + 2X P_{,XX}},
\end{equation}
and a comma denotes partial differentiation. 
The corresponding spectral indices are given by 
\begin{equation}
\label{spectral}
1-n_s = 2\epsilon + \frac{\dot{\epsilon}}{\epsilon H} + 
\frac{\dot{c}_s}{c_s H} , \quad 
n_t =-2 \epsilon
\end{equation}
and the tensor-scalar ratio, $r \equiv \Pt /\Ps$, is 
\begin{equation}
\label{conseqn}
r = 16 c_s \epsilon.
\end{equation}
In canonical inflation 
the sound speed of inflaton fluctuations 
is equal to the speed of light, $c_s=1$. The scalar spectral 
index is therefore determined by 
\begin{equation}
1-n_s = 6 \epsilon - 2\eta,
\end{equation}
where 
\begin{equation}
\epsilon=\frac{\mpl^2}{2}\(\frac{V'}{V}\)^2 , \quad 
\eta=\mpl^2\frac{V''}{V}
\end{equation}
represent the `slow-roll' parameters and 
a prime denotes $d/d\phi$. 
Successful slow-roll inflation requires a sufficiently 
flat potential, $\{ \epsilon , |  \eta | \} \ll 1$.

The observed normalisation of the CMB anisotropy power spectrum implies that 
$\Ps = 2.5 \times 10^{-9}$. If it is assumed that 
the scalar spectral index is effectively constant over observable scales
and that $r$ is negligible, the WMAP5 data \cite{wmap5} indicates that 
\be
\label{nbound5}
n_s=0.963^{+0.014}_{-0.015}
\ee
at the $1\sigma$ confidence level.  
Combining WMAP5 with the BAO and SN data 
reduces the uncertainty in the matter density and results
in an improved bound on $n_s$ when the tensor-to-scalar ratio
is included \cite{wmap5}. To be consistent we also use the bounds
from this data set in the case where $r$ is negligible:
\be\label{nbound}
n_s=0.96 ^{+0.014}_{-0.013}.
\ee
It is important to note that   
even if the tensor-scalar ratio is significant, the Harrison-Zeldovich 
spectrum $n_s=1$ is still highly disfavoured. On the other 
hand, if the running of the spectral index, $n_s'$, is treated as a free 
parameter, $n_s = 1$ would be consistent with the data. 
The upper limit on the tensor-scalar ratio from the combined 
${\rm WMAP5} + {\rm BAO}+ {\rm SN}$ data is $r<0.25$ at the $2\sigma$ level
when $n_s'=0$ is assumed as a prior. This is relaxed to
$r<0.54$ for $n_s'\neq0$.  

Deviations from Gaussian statistics in the scalar perturbation spectrum 
are quantified in terms of the `non-linearity' parameter, $\fnl$. 
In the equilateral triangle limit, where the three momenta have equal 
magnitude, the leading-order contribution to this quantity is 
determined by the kinetic function and its first three derivatives
\cite{M2002,LS2005,Chen2006}: 
\begin{equation}
\label{deffnl}
\fnl = \frac{5}{81} \left( \frac{1}{c_s^2} -1 -2 \Lambda \right) -
\frac{35}{108} \left( \frac{1}{c_s^2} -1 \right),
\end{equation}
where
\begin{equation}
\label{defLambda}
\Lambda \equiv \frac{X^2P_{,XX}+\frac{2}{3} X^3P_{,XXX}}{XP_{,X}+
2X^2P_{,XX}}.
\end{equation}
Current limits on the non-linearity 
parameter are
$-151 < \fnl < 253$ at the $2\sigma$ level \cite{wmap5}.  

Finally, the number of e-folds of canonical, slow-roll
inflation that occurred 
from the time $t_*$ when observable scales first 
crossed the Hubble radius during inflation to the epoch $t_{\rm end}$ 
when inflation ended is given by 
\be
\label{N}
\caln=\frac{1}{\mpl^{2}} \int_{\phi_{\rm end}}^{\phi_*}\frac{V}{V'}d\phi.
\ee
The value of $\caln$ depends on the 
reheating temperature. Requiring baryogenesis to take place at or 
above the electroweak scale implies that $\caln \gtrsim 30$. 
A value of $\caln \simeq 60$ corresponds
to a GUT scale reheating and hence an instantaneous change from
inflation to relativistic matter domination. 

\section{Canonical Inflation}

In order to decide whether a particular model is favoured by the 
data or not we implicitly make the following assumptions in this Section:

\begin{enumerate}
 \item The curvature perturbation was sourced entirely from quantum 
fluctuations in a single, slowly rolling 
scalar inflaton field and is purely adiabatic. 
We work to leading-order in the slow-roll approximation.  
\item The running in the spectral index 
vanishes\footnote{This is a reasonable prior since the latest WMAP analysis 
concludes that there is no support from the data 
to treat the running as a free parameter \cite{wmap5}.}, $n_s' =0$.
\item The form of the potential under study remains valid until
the end of inflation.
\item The universe reheats instantaneously immediately 
after the end of inflation.
\item Assuming instantaneous reheating, 
a reasonable range of values for the number of e-folds between 
$t_*$ and  $t_{\rm end}$ is taken to be $\caln=54\pm7$,
in accordance with the literature and more specifically 
Ref.~\cite{Liddle:2003as}. (We note, however, that the number
of e-folds could lie outside this region in either direction if 
further assumptions are made). 

\end{enumerate}

In the following, therefore, 
when a model is said to be `excluded convincingly 
at more than a given confidence limit for a given value of $\caln$', 
this should be understood within the context of the above assumptions.

\subsection{Small-Field Canonical Inflation}

Small-field models are defined as those for which the 
variation in the inflaton field is less than the reduced Planck mass,  
$|\Delta\phi |\equiv  |\phi_*-\phi_{\rm end}|<\mpl$. Typically, 
the amplitude of gravitational waves
generated in such models 
is undetectably small and the spectral index therefore 
provides the key observational discriminator. 

A general form for the small-field inflaton potential is given by 
\be\label{new_modular_mutant}
V=V_0\left[1-\(\frac{\phi}{\mu}\)^p\right],
\ee
where $\{V_0 , \mu , p\}$ are constants. When $p$ is an integer 
and greater than 2, such a potential may be generated 
by the self-coupling of the inflaton at tree-level. A potential 
of this form also arises in the new 
\cite{Linde:1981mu,Albrecht:1982wi,Banks:1995dp} and modular inflationary
scenarios 
\cite{Linde:1984cd,Binetruy:1986ss}. 
The mutated hybrid models correspond to the ranges 
$2< p< \infty$ or $-\infty < p < 1$, where $p$ is not necessarily an integer
\cite{mutant_stewart}. 
The case $p=-4$ corresponds to certain braneworld models, 
where the constant term arises from the warped tension of the brane and
anti-brane and the interaction term originates from the attraction between 
the branes \cite{DT,KKLT}. 

Another small-field model of interest  
is the supersymmetric (SUSY) $F$-term potential 
resulting from one-loop corrections of the 
waterfall field in SUSY hybrid inflation
\cite{Copeland:1994vg,Dvali:1994ms,Stewart:1994ts}. In this scenario, 
supersymmetry is broken spontaneously and  
the potential has the form
\be\label{log}
V=V_0\left(1+\frac{g^2}{8\pi^2}\ln\frac{\phi}{Q}\),
\ee
where $Q$ and $g$ are model-dependent parameters.

Finally, we consider the potential 
\be\label{exp}
V=V_0\(1-e^{-q\phi/\mpl}\),
\ee
with $q=\sqrt{2}$, which may be generated when the flat direction in 
SUSY is lifted by adding a Kahler potential, i.e., by adding kinetic terms
\cite{Stewart:1994ts}. (Note, however, that this is not a small-field 
model as defined above, 
since $\phi > \mpl$ in the region where $V_0$ dominates.)

All the above models are related through their observational predictions 
for the value of the spectral index.  
For both the potentials (\ref{log}) and (\ref{exp}), 
the slow-roll parameter $\eta$ dominates over $\epsilon$ during inflation.  
However, this is only true for the potential (\ref{new_modular_mutant})
in the small-field regime and we therefore implicitly 
assume\footnote{One could consider cases
where $\mu>\mpl$, but this would correspond to a large-field model, 
and would result in a larger tensor fraction for which \eq{np} would not apply 
\cite{Martin:2006rs}.} that $\mu<\mpl$. 
The fact that $\eta \gg \epsilon$ implies that 
the spectral index is given by $n_s-1 \simeq 2\eta$ and, 
since $\epsilon$ is a 
continuously increasing function of $\phi$ during inflation, 
this expression can be combined 
with the number of e-folds, Eq. (\ref{N}), to deduce a general form 
for the spectral index \cite{Kinney:1995cc,Liddle:2000cg}: 
\be
\label{np}
n_s=1- 2\(\frac{p-1}{p-2}\)\frac{1}{\caln}.
\ee
In Eq. (\ref{np}), 
the SUSY F-term hybrid model (\ref{log}) corresponds formally to 
$p=0$ and the potential (\ref{exp}) to the value $p= -\infty$. 

\begin{figure}
\includegraphics[width=\linewidth]{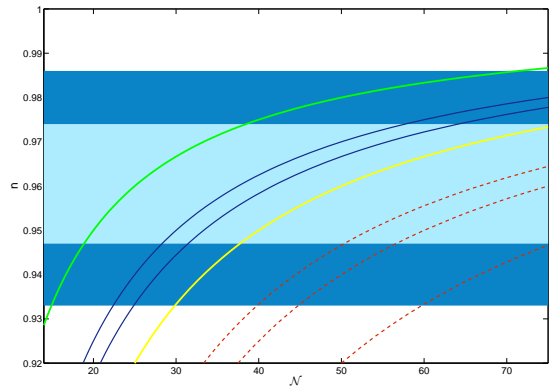}
\caption{Illustrating the dependence according to 
Eq. (\ref{np}) of the spectral index, $n_s$, on 
the number of e-folds, $\caln$, for fixed values of $p$. The 
light blue (dark blue) regions represent the 
$1\sigma$ $(2\sigma )$ confidence limits, respectively. 
The green line (uppermost line) is the model (\ref{log}) with $p=0$ and  
the yellow line corresponds to the potential (\ref{exp}) with 
$p=-\infty$. 
The blue lines represent mutated hybrid inflation with $p=-3,-4$.
The red lines are modular 
inflation models with $p=3,4,5$, where $p=3$ corresponds to the lowest 
line. 
}
\label{nvN5BAO}
\end{figure}

\begin{figure}
\includegraphics[width=\linewidth]{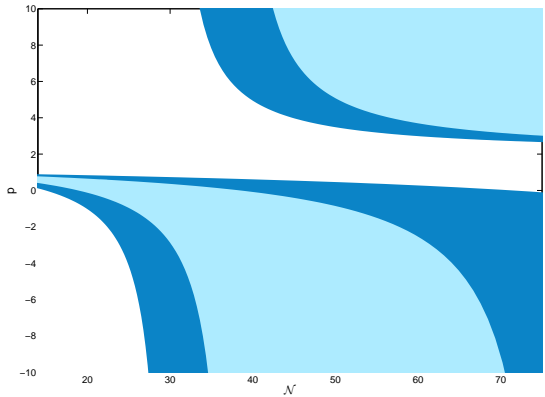}
\caption{Illustrating the power $p$ in Eq. (\ref{np}) 
versus the number of e-folds for fixed values of the spectral 
index. The light blue (dark blue) regions represent the 
$1\sigma$ $(2\sigma )$ confidence limits, respectively.}
\label{pvN5BAO}
\end{figure}

\begin{figure}
\includegraphics[width=\linewidth, totalheight=2.5in]{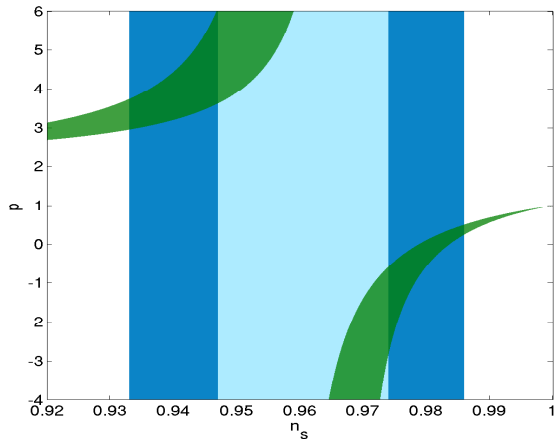}
\caption{The dependence of the power, $p$, on the 
spectral index, $n_s$, according to Eq. (\ref{np}). 
The green shaded regions
correspond to the range of e-folds $\caln=54\pm7$. 
The overlap between the green and blue areas 
yields the values of $p$ consistent with the observations
for this range of $\caln$.}
\label{pvn5BAO}
\end{figure}

In Figs. \ref{nvN5BAO}-\ref{pvn5BAO}, we explore the regions of 
parameter space defined by $\{ n_s ,p, {\cal{N}} \}$ that are 
consistent with the observations. In all cases, the bound (\ref{nbound}) 
on the spectral index is employed. 
In Fig. \ref{nvN5BAO}, 
the number of e-folds, $\caln$, is treated as an independent variable 
and the spectral index, $n_s$, is varied. The relation (\ref{np}) is plotted 
for a range of values for $p$. In Fig. \ref{pvN5BAO}, the parameter $p$ is varied 
with respect to $\caln$ for a fixed $n_s$. The bounds (\ref{nbound}) 
are highlighted to emphasise the allowed regions of parameter space. 
Finally, in Fig. \ref{pvn5BAO}, the value of $p$ is varied with respect to 
the spectral index for fixed values of the number of e-folds in the 
range $\caln=54\pm7$. 

\begin{table}[ht]
\centering
\begin{tabular}{|c|c|c|c|}
\hline
&Outside the&Outside the\\
&$1\sigma$ region&$2\sigma$ region\\
\hline
$\caln=47$&$p<6.1$&$p<3.74$\\
\hline
$\caln=54$&$p<4.32$&$p<3.23$\\
\hline
$\caln=61$&$p<3.62$&$p<2.96$\\
\hline
\end{tabular}
\caption{The exclusion limits for positive values of $p$ for particular 
values of $\caln$, 
based on the combined ${\rm WMAP5}+{\rm BAO}+{\rm SN}$ data on the 
spectral index. All values of 
$p<0$ are included at the $1$ and $2\sigma$ levels.}
\label{exclusion}
\end{table}

In \cite{Alabidi:2006qa}, the original hybrid inflation 
model was excluded convincingly at more than $3\sigma$. 
This is still the case with the 
new data sets. On the other hand, the $p<0$ (mutated hybrid) and 
$|p|\to\infty$ (exponential) models fit the data 
well for the reasonable range of $\caln$. 
Within the context of the present discussion, 
the key features of the new data are the increase 
in the central value of the spectral index and the narrowing 
of the error bars. As a result, the 
$p=3$ model of 
the new and modular inflationary scenarios 
is now outside the $2\sigma$ region for $\caln<60$, 
as can be seen from Fig. \ref{nvN5BAO}. Referring to 
\fig{pvn5BAO} and considering positive values of $p$, 
we find that $p > 3.74$ is required for inclusion at the $2\sigma$ 
level if $\caln=47$, whereas $p> 2.96$ is necessary if $\caln=61$. 
Moreover, for inclusion at $1\sigma$, these limits strengthen to  
$p > 6.1$ if $\caln=47$ and $p> 3.62$ if $\caln=61$. 
The exclusion limits for various values of $\caln$ are 
summarised in Table \ref{exclusion}. 

The fact that it is the lower-order terms
that are constrained by the data 
is important from the theoretical point of view,  
because it is difficult to motivate the suppression 
of lower-order terms in a small-field theory. 
Such a suppression requires either a significant fine-tuning of the 
coupling terms or the introduction of a suitable symmetry. 
The problem of suppressing the quadratic term in 
modular inflation was addressed in Ref. \cite{Ross:1995dq}, 
where a symmetry was identified. However, the majority of model
builders prefer to impose fine-tuning  
to suppress this term \cite{Lalak:2005hr} and  
this emphasises the difficulty of constructing a small-field model of 
inflation where a higher-order term is responsible for inflation 
coming to an end. This is an important 
open question that needs to be addressed. 

\subsection{Large-Field Canonical Inflation}

Large-field models are characterised by the condition 
$|\Delta \phi | \gtrsim \mpl$. They are particularly interesting since  
a super-Planckian field variation may generate an observable 
tensor-scalar ratio
\cite{Lyth:1996im}. The simplest class of large-field models is 
based on a monomial potential, $V \propto \phi^{\alpha}$,
where $\alpha$ is a positive \cite{Linde:1983gd} or negative 
\cite{Barrow:1990vx,Barrow:1990td,Muslimov:1990be}
constant. The former case corresponds to chaotic inflation,
whereas the latter corresponds to the 
intermediate inflationary scenario which was analysed in 
light of the WMAP3 data in Ref. \cite{Barrow:2006dh}.  
Assuming that the potential maintains this form up to the end of inflation, 
the generic predictions for this class of models are:
 
\begin{eqnarray}
\label{extra1}
1-n_s = \frac{2+\alpha}{2\caln}, 
\quad r= \frac{8 [ {\cal{N}} (1-n_s) -1 ]}{{\cal{N}}}
\end{eqnarray}
for $\alpha>0$, and
\begin{eqnarray}
\label{extra2}
1-n_s=\frac{|\alpha|(|\alpha|-2)}{\phi^2_*}, 
\quad r=\frac{8\alpha^2}{\phi^2_*}
\end{eqnarray}
for $\alpha<0$.

\begin{figure*}
\includegraphics[width=\linewidth]{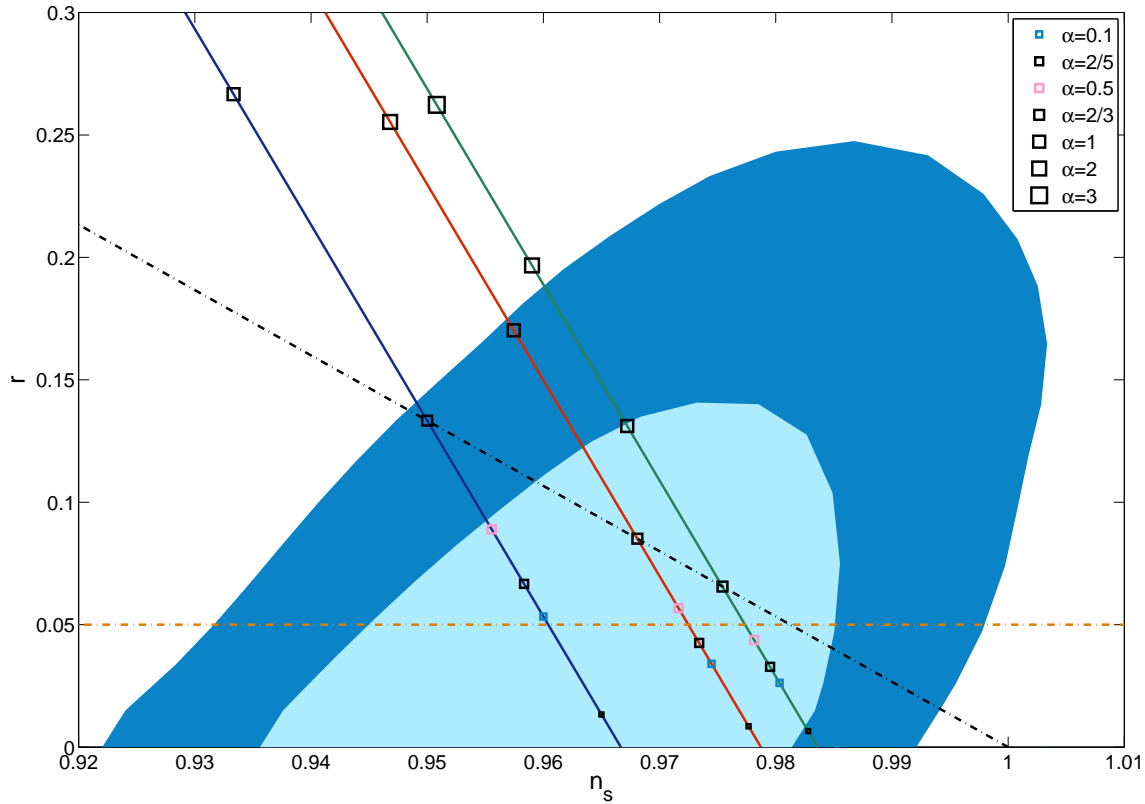}
\caption{Illustrating the dependence of the tensor-scalar ratio, $r$, 
on the spectral index, $n_s$, for a given number of e-folds, $\caln$, 
when the inflaton potential has a monomial form with $\alpha >0$. 
The light blue (dark blue) regions represent the $68\%$ ($95\%$) confidence
limits, respectively.
The blue line corresponds to $\caln=30$, the red line to $\caln=47$ 
and the green line to $\caln=61$. The black, dash-dotted line represents 
the $\alpha =1$ ($\eta=0$) linear potential, below which potentials are 
concave downward. 
The observational data is from the combined 
${\rm WMAP5} +{\rm BAO}+{\rm SN}$ set applied to the $\Lambda {\rm CDM}+
{\rm tensor}$ model without running in the spectral index. 
The contours were generated using the Matlab scripts provided
by the Cosmomc package and are smoothed to one degree. 
The horizontal dashed line is the expected sensitivity of 
the Planck satellite for detecting the tensor-scalar 
ratio. 
}
\label{power}
\end{figure*}

\begin{figure}
 \includegraphics[width=\linewidth]{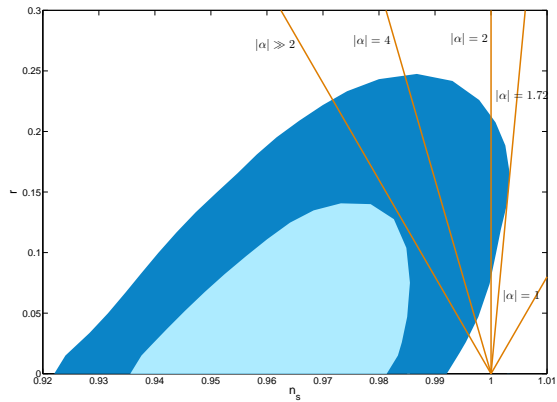}
\caption{The predictions of the intermediate 
inflationary scenario driven by a monomial potential with $\alpha<0$. 
The shaded regions are defined in Fig. \ref{power}.}
\label{inter}
\end{figure}

Since the predictions of a particular model are 
sensitive to the number of e-folds that elapsed between observable scales 
leaving the horizon and the end of inflation, 
we have plotted Eq. (\ref{extra1})  
for different values of $\caln=(30,47,61)$ in Fig. \ref{power}. 
On each line we have identified the predictions for specific 
values of $\alpha$ in the range $0.1\le \alpha \le 4$. 
We note that for potentials with $\alpha\geq2$, 
$\caln>61$ is required in order for the
predictions to lie inside the $1\sigma$ contour. 
 
It is interesting to consider what one might be able to deduce
from the $(r,n_s)$ plane in the near future with 
regard to concave-downwards potentials $(\alpha < 1)$, especially given the  
anticipated results from the Planck surveyor. This class of models 
is of relevance to braneworld inflation 
within the context of type IIA string theory compactified on 
a six-dimensional, bulk space that is comprised of  
two nil-manifolds \cite{Silverstein:2008sg}. 
Depending on the direction in which the brane moves, 
the leading-order term in the effective inflaton 
potential is a monomial with $\alpha=2/3$ or $\alpha =2/5$. 

The Planck surveyor is expected 
to be sensitive to a primordial gravitational wave background 
of $r \gtrsim 0.05$, as shown by the horizontal dashed line in 
Fig. \ref{power}.  For illustrative purposes, let us suppose  
that the allowed range of values for $n_s$ 
deduced from the Planck data at a given confidence level 
turns out to be similar to 
that from WMAP5 with the same set of 
priors\footnote{Naturally, we do not expect it 
to be the same. The purpose of this discussion is to illustrate 
how the method could be employed in practice when the data is made 
available.}. In that case, Fig. \ref{power} indicates that 
a tensor background would not be 
detectable for $\alpha \lesssim 2/3$ when the number of e-folds 
is in the range  
$\caln=54\pm7$, but could be for $\caln \simeq 30$, for example. The $\alpha
=2/5$ model, on the other hand, predicts 
that $r \lesssim 0.05$ for all $\caln \gtrsim 30$. 

Finally, we consider the implications of WMAP5 for the intermediate 
inflationary scenario. Rearanging Eq. ({\ref{extra2})  
implies that \cite{Barrow:2006dh}
\be\label{r_inter_n}
1-n_s = \frac{|\alpha | -2}{8|\alpha |} r
\ee
and Eq. (\ref{r_inter_n}) is plotted in Fig. \ref{inter} for various values of
$\alpha$. It should be noted that this model  
leads to eternal inflation and some hybrid mechanism is therefore required 
in order to bring inflation to an end. In this sense, 
the value of the inflaton field at the 
end of inflation is unspecified and, consequently, any value for 
$\caln$ is possible in principle. Nonetheless, we deduce that 
$r \rightarrow 8(1-n_s)$ in the limit $|\alpha | \gg 2$. This limit 
is shown in Fig. \ref{inter} and does not intersect the 1$\sigma$ 
region. For this particular combination of 
data sets, therefore, the intermediate scenario 
lies outside the $1\sigma$ contour for all values of $\alpha$ and  
all models with $|\alpha | < 1.72$ lie outside the $2\sigma$ regime.  

\section{Inflation with Wrapped Branes}

The DBI inflationary scenario is based on the 
compactification of type IIB string theory on a Calabi-Yau 
(CY) three-fold \cite{DBI}. Non-trivial form-field fluxes generate warped `throat' 
regions within the CY space and inflation arises when a 
D-brane propagates inside such a throat. The inflaton parametrises
the radial position of the brane and, 
since this is an open string mode, 
its dynamics is determined by a DBI action. 

In general, the metric inside the throat has the 
form $ds^2=h^2ds_4^2+h^{-2} ( d\rho^2 +\rho^2ds^2_{X_5})$, where 
the warpfactor $h(\rho)$ is a function of the radial coordinate $\rho$ 
along the throat and the base $X_5$ is an Einstein-Sasaki manifold. When
the throat geometry is $AdS_5 \times X_5$, 
the warpfactor is $h = \rho /L$, where $L^4 \equiv (4 \pi^4g_sN)/[\Vol m_s^4]$
defines the ${\rm AdS}$ radius, $N$ represents the ${\rm D3}$-brane charge
on the cone, $\Vol$ is the dimensionless 
volume of the base with unit radius, and $m_s$ and $g_s$ denote the 
string mass and coupling, respectively.  

A well-motivated exact warped geometry is provided by the 
Klebanov-Strassler (KS) background, where the base is $X_5=T^{1,1}$ with 
${\rm Vol}(T^{1,1}) = 16\pi^3/27$ \cite{KS}. Moreover, in the 
F-theory interpretation of type IIB compactifications, the 
global tadpole cancellation relates the background charge to 
the Euler characteristic, $\chi$, of the CY four-fold such that 
$N = \chi /24$ \cite{WITTEN}. For the case of four-fold manifolds 
that are known explicitly, 
$\chi < 1,820,448$, which implies that $N < 75,852$ \cite{KLEMM}. 

In the original version of the scenario, inflation was driven by the motion 
of a ${\rm D3}$-brane \cite{DBI}. However, this model is inconsistent 
with the data if
$| \fnl |$ is large \cite{lidseyhuston,peiris,bean,Lorenz}. 
Recently, extended versions involving 
${\rm D5}$- and ${\rm D7}$-branes wrapped over cycles of the throat
have been considered \cite{BECKER,KOB}. The key difference between these models 
is the relationship between the inflaton field and the radial coordinate.  
When the NS-NS two-form potential vanishes this is given by 
\begin{equation}
\label{keydiff}
\phi_{2n+3} = \sqrt{T_{2n+3} v_{2n} L^{2n}} \rho
\end{equation}
for an $AdS_5 \times X_5$ throat,
where we denote the ${\rm D}(3+2n)$-brane by the subscript $n = (0,1,2)$, 
$T_{3+2n} \equiv m_s^{2n+4}/[(2\pi)^{3+2n} g_s]$
defines the brane tension
and $v_{2n}$ represents the wrapped volume. (Note that  
$v_0 \equiv 1$.) 

The effective four-dimensional action for the brane 
dynamics is then given by Eq. (\ref{generalaction}) with a kinetic
function \cite{BECKER,KOB}: 
\begin{equation}
\label{Pwrapped}
P_{2n+3} = -T(\phi ) 
\left( 1- \frac{2X}{T(\phi )} \right)^{1/2} -V(\phi ) 
+ T(\phi  ) ,
\end{equation}
where the warped brane tension is defined by 
\begin{equation}
\label{warptension}
T(\phi_{2n+3}) \equiv T_{2n+3} v_{2n} L^{2n} h^4 (\phi_{2n+3} ).
\end{equation}
The warped nature of the geometry imposes an upper limit on the 
inflaton's kinetic energy. In this and the following Section,
we consider the `relativistic, ultra-violet' DBI scenario, where 
this upper limit is saturated, $\dot{\phi}^2 \simeq T$, and the 
brane is moving towards the tip of the throat, $\dot{\phi} <0$. 
We do not specify the form of the inflaton potential 
(subject to the condition that a phase of quasi-exponential expansion
is generated). It follows that  
\begin{eqnarray} 
\label{kinsat}
\dot{\phi}_{2n+3} \simeq - A_{2n+3} \phi^2_{2n+3},
\nonumber \\ 
A_{2n+3}^2 \equiv \frac{1}{T_{2n+3}v_{2n}L^{2n+4}}.
\end{eqnarray}
Moreover, the scalar perturbation amplitude has the form
\begin{equation}
\label{scalaramp}
\Ps = \frac{H^4}{4\pi^2\dot{\phi}^2} 
\end{equation}
and, since $c_s \ll 1$, 
the non-linearity parameter is given by $\fnl = - 1/(3c_s^2)$
\cite{Chen2006,lidser2}. 

The observable parameters in this scenario 
are $\{ \Ps, r, n_s , \fnl \}$, whereas the key parameters describing  
the warped throat are $\{ \Vol , g_s , N\}$. Our aim is to constrain 
these parameters with the WMAP5 data in a way that is independent 
of the inflaton potential. We proceed by 
substituting Eq. (\ref{kinsat}) into the scalar perturbation amplitude to 
deduce that 
\begin{equation}
\label{inflatonHubbleratio}
\left( \frac{H}{\phi_{2n+3}} \right)^4 = 4\pi^2\Ps A^2_{2n+3}.
\end{equation}
Assuming a quasi-de Sitter expansion, we differentiate Eq. 
(\ref{inflatonHubbleratio}) with respect to the comoving wavenumber, $k=aH/c_s$
\cite{GM1999,peiris,bean}. Substituting 
Eqs. (\ref{conseqn}) and (\ref{inflatonHubbleratio}) into the result
then yields a constraint equation involving the spectral index: 
\begin{equation}
\label{constraint1}
r \sqrt{-\fnl \mathcal{P}_S} - \frac{4}{\sqrt{3}} (1-n_s)
\sqrt{\mathcal{P}_S} = Q,
\end{equation}
where we have defined the quantity 
\begin{eqnarray}
\label{defQ}
Q & = & \frac{16}{\sqrt{3}}
\left( \frac{A^2_{2n+3}}{4\pi^2} \right)^{1/4}
\nonumber \\
& = & \frac{1}{\sqrt{3}} \left[ \frac{2^{15+n}}{\pi^3 v_{2n}}
\frac{1}{g_s^{n/2}} \left( \frac{\Vol}{N} \right)^{(2+n)/2} \right]^{1/4}.
\end{eqnarray}

The left-hand side of Eq. (\ref{constraint1}) 
is determined entirely by observable quantities, whereas the right-hand 
side depends on field-theoretic parameters. 
Since WMAP5 strongly favours a red perturbation spectrum ($n_s <1$) \cite{wmap5}, 
Eq. (\ref{constraint1}) yields the upper limit 
\begin{equation}
\label{upperlimitQ}
Q < r \sqrt{- \fnl \mathcal{P}_S} < 0.02 
\end{equation}
once current observational bounds are imposed.  
It follows, therefore, that rearranging Eq. (\ref{defQ}) in the form  
\begin{equation}
\label{vol/N}
\frac{\Vol}{N} = \left[ \frac{9\pi^3v_{2n}}{2^{15+n}} Q^4 
\right]^{2/(2+n)} g_s^{n/(n+2)}
\end{equation}
leads to an observational upper limit on the ratio $\Vol /N$. 
Invoking the canonical values $v_2 \simeq 4 \pi$ and 
$v_4 \simeq 8\pi^2/3$ for the wrapped volumes, respectively, we find that 
\begin{eqnarray}
\label{obsconstraint}
\left. \frac{\Vol}{N} \right|_{\rm D5} 
< 4.2 \times 10^{-6} g_s^{1/3} \nonumber \\
\left. \frac{\Vol}{N} \right|_{\rm D7} 
< 9.5 \times 10^{-5} g_s^{1/2}.
\end{eqnarray}

With a typical value of $\Vol \simeq {\cal{O}} (\pi^3 )$, the ratio
$\Vol /N \gtrsim 4 \times 10^{-4}$ for all currently known CY four-folds. 
We may conclude, therefore, that consistency 
with the data for both the wrapped ${\rm D}$5- and ${\rm D}$7-brane 
models would require either a smaller than expected 
value for the base volume or the discovery of new classes of CY 
four-folds with a larger Euler number. 

The theoretic lower bound on $\Vol /N$ can be reduced 
by allowing the brane to wind around the cycle of the throat
or by wrapping it around a base that is orbifolded. For example, if a 
${\rm D5}$-brane has a winding number $p$ and wraps around a cycle 
of $T^{1,1}$ that is orbifolded by $Z_q$, the normalisation  
$S \equiv \phi^2/\rho^2$ is altered by a factor of $p/q$ \cite{BECKER}. 
This modifies the value of the parameter $A_5$ defined in Eq. (\ref{kinsat}) 
by a factor $(q/p)^{1/2}$ and, consequently, the value of $Q$ by 
$(q/p)^{1/4}$. It follows, therefore, that 
the upper bound (\ref{obsconstraint}) for the ${\rm D5}$ scenario becomes 
\begin{equation}
\label{orbifold}
\left. \frac{\Vol}{N} \right|_{\rm D5} < 4.2 \times 10^{-6} 
\left( \frac{p}{q} \right)^{2/3} g_s^{1/3}.
\end{equation}
This implies that a large amount of winding or orbifolding 
is required. 

It is interesting to consider how the constraints (\ref{obsconstraint}) 
compare to previously known bounds in the light of the WMAP5 data. 
There exists a field-theoretic upper bound on the tensor-scalar ratio 
in these scenarios, which arises because 
the warped nature of the bulk manifold 
restricts the maximally allowed variation in the value of the inflaton field 
\cite{bmpaper,lidseyhuston}. For the ${\rm D}5$ and ${\rm D7}$ models, 
this constraint was recently expressed in terms of the upper bound 
\cite{KOB}  
\begin{equation}
\label{KMKlimit}
\left( \frac{\Vol}{N} \right)^{n+2} < 
4^{(1-n)} \pi^{6} g_s^{n}
v_{2n}^{2} \mathcal{P}_S^{4}.
\end{equation}
Comparison with Eq. (\ref{defQ}) implies that this constraint 
can be expressed succinctly in the form 
\begin{equation}
\label{MKMsuccinct}
Q < 16  \left( \frac{\mathcal{P}_S}{3} \right)^{1/2}.
\end{equation}

The WMAP3 data imposed the limits $\fnl < -300$ and $r <0.5$ at the $1\sigma$
level, which implies that the upper bound (\ref{upperlimitQ}) 
would have been $Q < 0.06$. 
This is comparable in strength to the theoretic 
bound (\ref{MKMsuccinct}). Indeed, prior to the WMAP5 data, 
the wrapped ${\rm D7}$-brane scenario was still 
compatible with such a limit if $\Vol \simeq {\cal{O}} (\pi^3)$ 
and $g_s > 0.2$ \cite{KOB}. However, 
the WMAP5 data has strengthened the bound (\ref{obsconstraint}) 
on $\Vol /N$ by a factor of $3^{8/(n+2)}$ with the result that all 
wrapped versions of the scenario now 
face theoretical challenges if they are to be consistent with 
the observations. 

It is also interesting to note that post WMAP5, 
the limit (\ref{orbifold}) is comparable numerically 
to a corresponding limit for the ratio $\Vol /N$ derived in 
Ref. \cite{BECKER}. (The limit (\ref{orbifold}) 
would have been weaker by a factor of approximately $20$ if the WMAP3 data 
had been employed). 
However, the analysis of Ref. \cite{BECKER} was restricted to 
the specific case of a quadratic potential and did not consider the 
observational constraints on the spectral index. 
Since the above analysis was independent of the potential, 
this indicates that any modifications to $V(\phi )$ are 
unlikely to significantly weaken the upper limit on $\Vol /N$. 

In conclusion, therefore, one of the consequences of the WMAP5 data is that  
the strongest constraints on the wrapped, relativistic, UV DBI inflationary
scenario now arise from the observable $(r, n_s )$ plane. 
On the other hand, it is important to note 
that backreaction effects are more significant in the wrapped 
scenarios when the brane is moving relativistically \cite{BECKER,Bean:2007eh}. 
Consequently, a more detailed analysis of such effects is required before 
these models can be concretely excluded. Nonetheless, it  
is known that a wrapped ${\rm D}5$-brane with flux is dual to $n$ coincident 
${\rm D}3$-branes in the large $n$ limit \cite{WARD}. In view of this, we 
proceed in the next Section to consider inflation from multiple-brane 
configurations.
 
\section{Inflation with Multiple Branes}

In general, the theory of multiple-brane configurations is non-Abelian since 
$n$ coincident branes have open
string degrees of freedom which combine to fill out representations of
$U(n)$. Recently, the effective action for 
$n$ coincident ${\rm D}3$-branes was derived in the finite $n$ limit
by employing an iterative technique based on the fundamental 
representation of $SU(2)$ and taking an appropriate symmetrized trace over the
gauge group \cite{thomasward,HLTW}. It was found that for all values of $n$, the
kinetic function has the form 
\begin{eqnarray}
\label{finiteP}
P= 2 T_3 \left\{ h^4 \sqrt{1+(n-1)^2 Y} \left( 1- \frac{\dot{\phi}^2}{T_3h^4}
\right)^{-1/2} \right\} \nonumber \\
-nT_3 (V-h^4) 
\end{eqnarray}
in the `relativistic' limit $\dot{\phi}^2 \simeq T_3h^4$, where 
$Y$ is defined by 
\begin{equation}
\label{defY}
Y \equiv \frac{1}{\pi^2(n-1)^4} \frac{m_s^4}{T_3^2} \left( \frac{\phi}{h}
\right)^4.
\end{equation}
It was further shown 
that backreaction effects are kept under control for sufficiently small 
values of $n \le 10$ \cite{HLTW}. 

We consider the relativistic, UV DBI scenario, 
where the branes are moving down  
an $AdS_5 \times X_5$ throat such that  
$h = \phi /(\sqrt{T_3} L)$ and $\dot{\phi} = -\phi^2/\sqrt{T_3L^4}$. 
The parameter $Y =  4\pi^2g_sN/[(n-1)^4 \Vol]$ therefore 
takes a constant value.
It follows from Eqs. (\ref{defcs}) and (\ref{finiteP}) that 
the non-linearity parameter is related to the sound speed 
of the inflaton fluctuations by $\fnl \simeq - 0.3/c_s^2$ \cite{HLTW}. 
It can be further shown that the scalar 
perturbation amplitude for this multi-brane scenario is given by \cite{HLTW}
\begin{equation}
\label{finitescalaramp}
\Ps = \frac{1}{50} \frac{H^4}{T_3h^4\sqrt{1+(n-1)^2Y}} \frac{1}{(-\fnl )}.
\end{equation}

In principle, this scenario could be constrained observationally by deriving 
the analogue of Eq. (\ref{constraint1}).  
However, the dependence of the scalar perturbation amplitude 
on the non-linearity parameter introduces an 
additional term involving the running $n_{NL} \equiv d \ln \fnl /d \ln k$. 
Although this parameter might be detectable in future experiments, 
there is presently no data available to constrain it. 
To proceed, therefore, we need to invoke a further assumption by 
specifying the form of the potential. In general, 
a mass term is expected to be generated 
below a critical scale due to the breaking of conformal invariance. 
We therefore consider a quadratic potential,  
$V=m^2\phi^2 /2$. 

To generate a phase of quasi-exponential expansion, 
$\epsilon = -\dot{H}/H^2 \ll 1$, the inflaton potential must dominate the 
energy density. In this case, the Friedmann equation reduces to 
$H^2 \simeq V/(3\mpl^2 )$ and it follows that 
\begin{equation}
\label{finiteepsilon}
\epsilon = \frac{1}{\sqrt{T_3L^4}} \frac{\phi}{H}  = {\rm constant}.
\end{equation}
The consistency equation (\ref{conseqn}) then implies that 
the scalar perturbation amplitude varies as 
$\Ps \propto 1/\fnl \propto c_s^2 \propto r^2
= \calp_T^4 /\calp_S^4$. Differentiating with respect to comoving 
wavenumber therefore implies that 
\begin{eqnarray}
\label{finiteconseqn}
\epsilon = \frac{3}{4} (1-n_s),
\nonumber \\
r = \frac{6.6(1-n_s)}{\sqrt{-\fnl}}
\end{eqnarray}
and we may deduce immediately that this model predicts a red 
perturbation spectrum, as favoured by the WMAP5 data. 

The ratio $\Vol /N$ can be constrained by 
substituting Eqs. (\ref{defY}) and (\ref{finiteepsilon}) into 
Eq. (\ref{finitescalaramp}). We find that 
\begin{equation}
\label{finitescalaramp1}
\Ps = - \frac{n-1}{50\pi^2 g_s^{1/2} \fnl \epsilon^4} 
\left( \frac{\Vol}{N} \right)^{3/2},
\end{equation}
and it follows from the consistency equation 
(\ref{finiteconseqn}) that 
\begin{equation}
\label{finiteratio}
\frac{\Vol}{N} = 29 (- \fnl )^{2/3} \calp^{4/3}_S (1-n_s)^{8/3} 
\frac{g_s^{1/3}}{(n-1)^{2/3}}.
\end{equation}
Imposing current WMAP5 upper limits on the observable parameters 
then implies that 
\begin{equation}
\frac{\Vol}{N} < 1.8 \times 10^{-6} \frac{g_s^{1/3}}{(n-1)^{2/3}}.
\end{equation}
We conclude, therefore, that when confronted 
with the observations, the multi-brane scenario 
(with a quadratic inflaton potential) faces the same 
theoretical challenges as the single, wrapped configurations: 
either a new class of CY four-fold is required  
or the base manifold $X_5$ must be orbifolded. 

On the other hand, despite such issues, it is interesting to note that 
the multi-brane scenario predicts a strong lower 
limit on the tensor-scalar ratio. Since $0.935<n_s<0.98$ at the 1$\sigma$ 
level, Eq. (\ref{finiteconseqn}) implies that $0.01 < r < 
0.43/\sqrt{-\fnl}$, where we have saturated the WMAP5 bound 
$\fnl > -151$ in the lower limit. 
This implies that the prospects for detecting primordial gravitational waves
from this model with future space-based, all-sky CMB polarisation 
experiments are good. Moreover, 
Eq. (\ref{conseqn}) may be expressed in terms of observable 
parameters such that $r+8n_t \simeq -\sqrt{-3\fnl} r$. 
As shown in Ref. \cite{lidser2}, 
deviations of this form from the standard 
inflationary consistency equation, $r= -8n_t$, 
could be detected in future CMB surveys if 
$r \gtrsim 0.2/\sqrt{-\fnl}$. Substituting this limit into 
Eq. (\ref{finiteconseqn}) therefore implies that such a detection could be 
possible if $n_s < 0.97$. This would provide a 
powerful observational discriminant for breaking the degeneracy 
between canonical and non-canonical, single-field inflation. 

\section{Discussion}

In this paper, we have considered the status of a wide class of 
single-field inflationary models in the light of the recent WMAP5 data. 
For the reasonable range of e-fold values aforementioned, 
$\caln = 54 \pm 7$, we conclude that 
modular/new inflation models
with $2<p\lesssim3$ are under pressure from the data 
and that $p<0$ and $|p|\to\infty$ satisfy the observations. 
The required suppression of lower-order terms in the $p>0$ models poses
a challenge to model builders, as discussed in Section III.
Moreover, it is significant that independent  
data sets are now converging towards a preferred 
value of $n_s$, with only tiny differences between  
WMAP5 only, ${\rm WMAP5}+{\rm SDSS}$ \cite{Percival:2006gt} and ${\rm WMAP5}+{\rm BAO}
+{\rm SN}$ data. This convergence implies that we can be more confident 
as to what represents a viable model from a phenomenological point of view. 

We have also shown that the intermediate inflationary scenario 
can only satisfy the data at $2\sigma$ for $|\alpha|\geq 1.72$, 
and that the predictions of the model do not overlap the $1\sigma$
region for any value of $\alpha$.

We emphasize that the above conclusions follow from assuming that the 
reheating of the universe immediately after inflation was instantaneous. 
However, the precise duration of the reheating process is not known. 
In particular, for a quadratic potential, 
a matter dominated phase may arise if the inflaton 
oscillates about its potential minimum for an extended interval before 
decaying. This would alter the number of e-folds
that elapsed from the epoch when observable scales first crossed 
the Hubble radius during inflation to the time when inflation ended. 
In general, the number of e-folds depends on the effective equation of state 
after inflation and this would need to be known if a more 
precise value of $\cal{N}$ is to be determined. 

Complementary to our work is that of Ref.~\cite{Peiris:2008be}, 
where the authors start from the different stand 
point of aiming to construct the most general inflationary potential using 
slow-roll reconstruction with a specific e-fold and reheat temperature prior. 
Reconstructing the inflationary slow-roll parameters and not 
the spectral parameters from the data has the advantage of 
directly probing more fundamental parameters. 
They locate regions in $\epsilon,\eta$
space which are consistent with the WMAP5 data, and this allows 
to reconstruct
the Hubble parameters and therefore (via the Hamilton-Jacobi equations) the 
inflationary potential. 
The authors also analyse higher-order parameters, but for the purposes of
this work, we only need mention that the simplest fit is consistent with the crucial
prior of $\caln>15$, which we interpret as support for the inflationary paradigm.

We have also considered the relativistic, 
ultra-violet, DBI braneworld scenario, where the 
inflaton is identified in terms of 
the radial position of a wrapped ${\rm D}5$- or 
${\rm D}7$-brane. We  found that when the brane's
kinetic energy is maximised, new constraints 
on the parameters of the bulk geometry can be derived from the 
WMAP5 data in the $(r , n_s )$ plane. Such constraints 
are independent of the precise form of the inflaton potential and are 
stronger than existing limits originating from field-theoretic bounds on the 
tensor-scalar ratio. In all cases, consistency with the data requires 
a significant reduction in the volume of the base or 
the discovery of new classes of Calabi-Yau four-folds
which allow for larger Euler numbers. 

We then extended our analysis to a recently proposed relativistic, 
multi-brane DBI scenario. For the case where the inflaton has a quadratic 
potential, we found that the predicted value of the spectral index 
is compatible with observations. However, the ratio  
$\Vol/N$ is still strongly bounded from above, as in the single brane 
models, and consequently the same theoretical problems arise in this 
case also. 

\section{Acknowledgements}
We thank D. Lyth and K. Malik for helpful discussions. We acknowledge use 
of the Cosmomc Matlab scripts. LA is supported by the 
Science and Technologies Facilities Council (STFC) under Grant 
PP/E001440/1.

\bibliographystyle{apsrev}
\bibliography{qmw08-1f}

\end{document}